\documentclass[12pt]{article}
\usepackage{amsmath,amssymb }
\usepackage{fancybox}
\usepackage{graphicx,psfrag,epsf}
\usepackage{enumerate}
\usepackage{graphicx,psfrag}
\usepackage{multirow}
\usepackage{epsfig}
\usepackage{subfigure}
\usepackage{theorem}
\usepackage{natbib,psfrag}
\usepackage[usenames,dvipsnames]{color}
\usepackage{todonotes}
\usepackage{xcolor}
\usepackage{algorithm2e}
\usepackage{url}
\usepackage{algorithm2e}

\usepackage{tabularx}
\usepackage{mathtools}
\newcommand{\pdf}{0}
\if1\pdf
\usepackage[pdftex]{graphicx}
\else
\usepackage{graphicx}
\fi

\newcommand{\blind}{0}

\def\mb#1{\setbox0=\hbox{$#1$}
	\kern-.025em\copy0\kern-\wd0
	\kern.05em\copy0\kern-\wd0
	\kern-.025em\raise.0em\box0}

\makeatletter
\newcommand*\dashline{\rotatebox[origin=c]{90}{$\dabar@\dabar@\dabar@$}}
\makeatother

\newtheorem{theorem}{Theorem}

\newtheorem{proposition}{Proposition}
\theoremstyle{definition}

\newtheorem{assumption}{Assumption}

\newcommand{\RN}[1]{%
	\textup{\uppercase\expandafter{\romannumeral#1}}%
}

\def\mb#1{\setbox0=\hbox{$#1$}
	\kern-.025em\copy0\kern-\wd0
	\kern.05em\copy0\kern-\wd0
	\kern-.025em\raise.0em\box0}
\def\0{{\bf 0}}


\newcommand{\sps}{\text{span}}
\def\mb#1{\setbox0=\hbox{$#1$}
	\kern-.025em\copy0\kern-\wd0
	\kern.05em\copy0\kern-\wd0
	\kern-.025em\raise.0em\box0}

\newcommand{\cor}{\text{Corr}}

\newcommand{\vect}{\text{vec}}

\begin{document}

\def\spacingset#1{\renewcommand{\baselinestretch}%
{#1}\small\normalsize} \spacingset{1}


\if0\blind
{
	\title{\bf A Cepstral Model for Efficient Spectral Analysis of Covariate-dependent Time Series}
	\author{Zeda Li  \hspace{.2cm}\\
		Paul H. Chook Department of Information System and Statistics,  Baruch College,\\ The City University of New York\\
		and \\
		Yuexiao Dong \\
		Department of Statistics, Operations, and Data Science,  Temple University}
	\maketitle
} \fi

\if1\blind
{
	\bigskip
	\bigskip
	\bigskip
	\begin{center}
		{\LARGE\bf Title}
	\end{center}
	\medskip
} \fi

\setlength{\baselineskip}{22pt}  
\begin{abstract}
	
This article introduces a novel and computationally fast model to study the association between covariates and power spectra of replicated time series. A random covariate-dependent Cram\'{e}r spectral representation and a semiparametric log-spectral model are used to quantify the association between the log-spectra and covariates. Each replicate-specific log-spectrum is represented by the cepstrum, inducing a cepstral-based multivariate linear model with the cepstral coefficients as the responses. By using only a small number of cepstral coefficients, the model parsimoniously captures frequency patterns of time series and saves a significant amount of computational time compared to existing methods. A two-stage estimation procedure is proposed. In the first stage, a Whittle likelihood-based approach is used to estimate the truncated replicate-specific cepstral coefficients. In the second stage, parameters of the cepstral-based multivariate linear model, and consequently the effect functions of covariates, are estimated.  The model is flexible in the sense that  it can accommodate various estimation methods for the multivariate linear model, depending on the application, domain knowledge, or characteristics of the covariates. Numerical studies confirm that the proposed method outperforms some existing methods despite its simplicity and shorter computational time. Supplementary materials for this article are available online.

\end{abstract}

\noindent%

\newpage
\spacingset{1.5} 
\section{Introduction}

Understanding the frequency domain properties of time series, as captured by the power spectrum, is crucial across various scientific disciplines. For instance, the frequency-domain analysis of biomedical time series, such as center-of-pressure trajectory, heart rate variability (HRV), and electroencephalography (EEG), offers valuable insights into the underlying physiological processes \citep{cabeza2011time, Halletal2004, Klimesch1999}. In many studies, researchers collect time series data from multiple subjects, referred to as replicated time series, to investigate the connections between these series and relevant covariates, such as behavioral or clinical outcomes. These analyses often involve examining the association between the power spectra and multiple covariates.

\begin{figure}[h]
	\centering
	\includegraphics[scale=0.6]{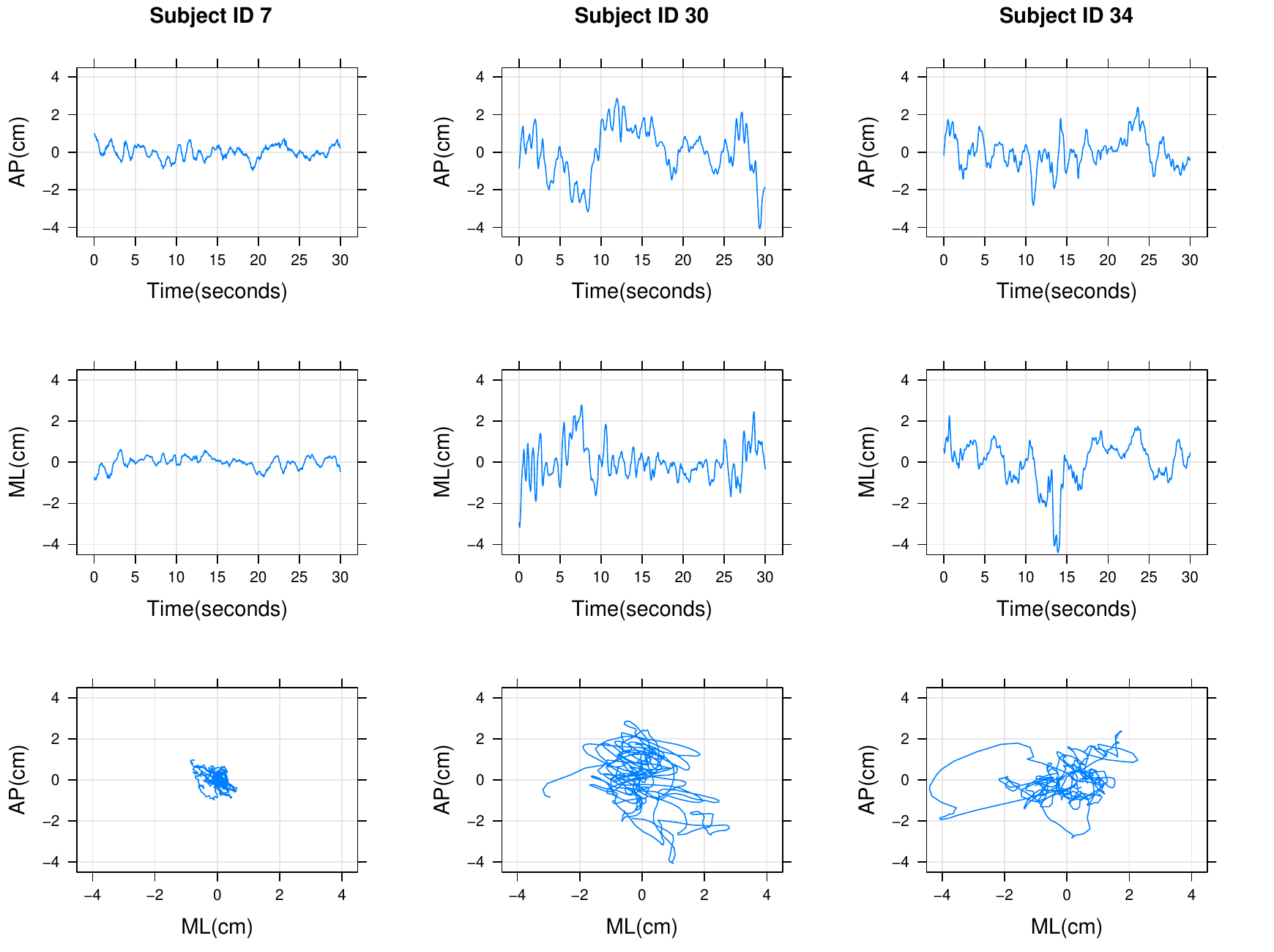}
	\caption{30 second COP trajectories sampled at 25 Hz for three patients standing with feet together and eyes closed. Each column corresponds to an individual patient. Rows (top to bottom) consist of the AP, ML, and joint AP and ML trajectories, respectively.}
	\label{realdata}
\end{figure}

A motivating application for this article comes from a study of human balance in Parkinson's disease (PD) patients. Impaired balance control is a significant symptom in PD, contributing to increased risks of falling and reduced quality of life \citep{grimbergenetal2013}.  The study seeks to enhance our understanding of the physiological mechanisms underlying postural regulation in PD patients and how postural instability correlates with clinical outcomes such as fear of falling, as assessed by scales like the Tinetti Falls Efficacy Scale (FES) and Activities-specific Balance Confidence Scale (ABC) \citep{tinettiFES, powell1995activities}. Postural instability was assessed by measuring changes over time in the center of pressure (COP) under the foot in both the anteroposterior (AP) and mediolateral (ML) directions while standing, commonly known as the COP trajectories. Of particular interest are the frequency patterns of COP trajectories, as characterized by the power spectrum, which provide objective physiological measures of postural instability and have been a major focus in the human balance literature over the last decades \citep{cabeza2011time}. Examples of COP trajectories for three different patients are displayed in Figure \ref{realdata}. By quantifying the relationship between COP power spectra and various covariates, this study aims to deepen our understanding of the link between postural instability and clinical outcomes.

Spectral analysis of replicated time series has garnered significant attention in recent years.  An extensive body of literature focused on quantifying the association between power spectrum and {\it qualitative} variables in experimental designs \citep{shumway1971, diggle1997, Fokianos2008, freyermuth2010tree, stoffer2010, chauandsachs2016, li2024anopow}. However, these methods are not applicable to incorporate {\it quantitative} covariates such as FES and ABC. When dealing with a {\it single} quantitative covariate, methods that can capture a smooth covariate effect on the power spectra  \citep{Fiecas2016, krafty2017,li2023robust} and methods that can adaptively capture both smooth and abrupt dynamics in power spectra across a covariate \citep{bruce2018,LiBruce2020MultiCABS} have been developed. Unfortunately, these methods, in their current form, cannot handle {\it multiple} quantitative covariates, which restrict their applicability to numerous important studies.

Existing methods that can account for {\it multiple quantitative} covariates include the semiparametric models of  \cite{iannaccone2001semiparametric} and \cite{krafty2011}. In these models, log periodograms are treated as responses, and the penalized least squares are used to obtain the smoothed effect functions. More recently, nonparametric Bayesian methods have been proposed \citep{Bertolacci2019, wang2023adaptive}. These methods rely on sophisticated Markov chain and Monte Carlo (MCMC) sampling schemes to fit the models. However, all aforementioned methods share two common limitations when dealing with complex and large time series datasets.  First, they either require smoothing parameters tuning through cross-validation or rely on complicated MCMC algorithms, making them computationally infeasible for datasets of moderate or large size.  Second, they don't provide sufficient means for dimension reduction or variable selection in the covariate space, which is much needed when a number of covariates are observed. Consequently, their applications to modern studies involving large and complex time series datasets are limited. The goal of this article is to introduce a flexible and computationally efficient framework for accurately quantifying the association between power spectra and multiple covariates, as well as providing means for dimension reduction or variable selections.

We propose a novel framework for analyzing replicated time series and a collection of covariates. The replicated time series and covariates are jointly represented by a random covariate-dependent  Cram\'{e}r spectral representation, and a semiparametric log-spectral model is employed to quantify the association between the log-spectra and covariates. Motivated by the successful use of cepstral coefficients in modeling log-spectra \citep{bogert1963, bloomfield1973,proietti2019generalized}, we represent the replicate-specific log-spectra using the replicate-specific cepstra. This formulation leads to a cepstral-based multivariate linear model with the replicate-specific cepstral coefficients of the log-spectra as the responses. By using only a small number of cepstral coefficients, the model parsimoniously captures frequency patterns of time series and saves a significant amount of computational time compared to the existing methods without sacrificing accuracy. We propose a two-stage approach for estimating the proposed model. In the first stage, we obtain the truncated replicate-specific cepstral coefficients through a Whittle likelihood-based approach, and the number of cepstral coefficients is determined by Akaike Information Criterion (AIC). In the second stage, we estimate the parameters of the cepstral-based multivariate linear model, and consequently, the effect functions of covariates. Our model is flexible in the sense that  it can accommodate various estimation methods for the multivariate linear model, depending on the application, domain knowledge, or characteristics of the covariates.


The remainder of the paper is organized as follows. Section 2 introduces key components of the proposed cepstral model, including the random covariate-dependent  Cram\'{e}r spectral representation, the semiparametric log-spectral model, the cepstral coefficients of replicate-specific time series, and the cepstral-based multivariate linear model.  Section 3 presents the proposed two-stage estimation procedure and explores the theoretical properties.  Simulation results are provided in Section 4. Section 5 presents an application to the study of postural control while standing in people with Parkinson's disease. Discussions and future directions of this work are covered in Section 6.  Additional simulation results and proofs are given in the Supplementary Materials.

\section{The Model}
\label{sec:model}

\subsection{Random Covariate-Dependent Cram\'{e}r Representation}
\label{sec:cramer}
We consider a collection of second-order stationary time series $\{Z_{jt},t=1, \dots, T\}$ and the corresponding $P$-dimensional vector of covariates $\mb X_j = (X_{j1}, \dots, X_{jP})^\intercal \in \mathcal{X}$ from  $j=1, \ldots, N$ independent subjects. We assume that the covariates $\mb X_j$  are independently and identically distributed with zero-mean and nonsingular covariance matrix $\mb \Sigma_{XX}$. We model the time series through a random covariate-dependent Cram\'{e}r spectral representation
\begin{equation*}\label{covariate_cramer}
Z_{j t} = \int_{-1/2}^{1/2} A_j(\mb X_j, \omega) \exp{(2 \pi i \omega t)}  d \Lambda_j(\omega),
\end{equation*}
where $A_j(\mb X_j,\omega)$ are the complex-valued random transfer functions defined over $\mathcal{X} \times \mathbb{R}$. For every $\mb X_j \in \mathcal{X}$, $A_j(\mb X_j, \cdot)$ is Hermitian, absolutely continuous, square-integrable over frequencies $[-1/2,1/2]$, and has period 1 as a function of frequency. Moreover, $A_j$ and $A_{j'}$ are independently and identically distributed (i.i.d.) conditional on $\mb X_j$ for $j, j' \ne 0$. $\Lambda_j(\omega)$ are mutually independent identically distributed mean-zero orthogonal increment processes over $[-1/2,1/2]$ such that $E\left\{  d \Lambda_{j}(\omega) d {\Lambda^*_{j}(\omega')} \right \}=d\omega$ if $\omega = \omega'$ and zero otherwise, where $*$ denotes the complex conjugate. In addition, $\Lambda_j( \omega)$  is independent of $A_{j'}(\mb X_{j'}, \omega)$ for all $j$ and $j'$. 
The time series $Z_{jt}$ exists with probability one, is short-range dependent,  and is mean-zero second-order stationary. To assure tractable estimation and inference, we assume the following regularity conditions
\begin{assumption}\label{assum_spectral}
~	\\
(a) $\lambda_{jt}= \int_{-1/2}^{1/2} \exp(2\pi i \omega t) d\Lambda_j(\omega)$ are white noise with mean zero and unit variance, and cumulants of $d\Lambda_j$ exist and are bounded for all orders \citep{brillinger2001}.\\
(b) There exists an integer $\rho > 0$ such that $\int|E(\exp(i s \lambda_{jt})|^\rho ds < \infty $.\\
(c)	$\sup_{\omega \in \mathbb{R}} E\left |A_j(\mb X_j, \omega)\right |^4 < \infty$, $\inf_{\omega \in \mathbb{R}} E\left |A_j(\mb X_j, \omega)\right |^4 > 0$, and there exists an $\eta>0$ such that 
$\sup_{\omega, \mb X_j} P\{|A_j(\mb X_j, \omega)|^2 < \eta\} = 0$.
\end{assumption}
Similar assumptions have been made in \cite{krafty2011}. Assumption \ref{assum_spectral}(a) also assures that $Z_{jt}$ can be written as a linear process. Assumption \ref{assum_spectral}(b) excludes $\lambda_{jt}$ with discrete distributions and is satisfied when $\lambda_{jt}$ possesses a differentiable density. Assumption \ref{assum_spectral}(c) ensures that the first two moments of $g_j(\mb X_j, \omega)$ exist and are bounded, and the $g_j(\mb X_j, \omega)$ are bounded away from zero. It should be noted that time series with discrete or a mixture of discrete and continuous power spectrum do not satisfy conditions in Assumption \ref{assum_spectral}, and thus, the cepstral representation considered in Section \ref{sec:cep}  may not be valid. However, similar assumptions are widely used in the spectral analysis literature \citep{brillinger2001,krafty2011,Fiecas2016}, and hold true for a wide variety of processes commonly encountered in practice, such as the autoregressive and moving average processes \citep{shumway2011}.

It is worth noting that, under this formulation, the covariates are not related to the mean of the replicated time series. Instead, they relates to the time series through the covariate-dependent transfer functions which capture how the second-order power spectra associate with the covariates. 
In particular, each time series with a fixed covariate $\mb X_j$ is i.i.d. mean-zero second-order stationary process possessing spectral density $E\{|A_j(\mb X_j, \omega)|^2\}$.  Conditional on the replicate-specific random transfer function $A_j$, $Z_{jt}$ is also mean-zero stationary and has replicate-specific power spectrum $|A_j(\mb X_j, \omega)|^2$. In many studies, such as our motivating example, scientific interest lies in the ratio of power at different frequencies. This is equivalent to examining linear combinations of the log power spectrum. Thus, we consider the replicate-specific log-spectrum $g_j(\mb X_j, \omega) = \log |A_j(\mb X_j, \omega) |^2$.  


\subsection{Semiparametric Log-spectral Model}\label{semi}

We propose a semiparametric model for the random transfer functions. Similar approaches have been considered in \cite{diggle1997}, \cite{iannaccone2001semiparametric} and  \cite{krafty2011}. Specifically, we express the random transfer functions as
\begin{equation*}
A_j(\mb X_j, \omega) =  \left [ h^{(0)}(\omega) \prod_{p=1}^P h^{(p)}(\omega)^{X_{jp}}\right]   h_j(\omega),
\end{equation*}
where $h^{(0)}(\omega)$ and $ h^{(p)}(\omega)$ are deterministic Hermitian functions with period 1 that are bounded away from zero, $h_j(\omega)$ are mutually independent Hermitian random functions with period 1 and $E\{h_j(\omega)\}=1$ for all $\omega$ that are also bounded away from zero, and $h_j$ and $h_{j'}$ are independently and identically distributed when $j \ne j'$. Define the functions $\alpha(\omega) = \log|h^{(0)}(\omega)|^2$, $\beta_p(\omega) = \log|h^{(p)}(\omega)|^2$ for $p=1,\cdots,P$, and $\xi_j(\omega) = \log|h_j(\omega)|^2$.  This formulation leads to a semiparametric model of the replicate-specific log-spectra and covariates
\begin{equation}\label{semilog}
g_j(\mb X_j, \omega) = \alpha(\omega) + \mb X_j^\intercal \mb \beta(\omega) + \xi_j(\omega),
\end{equation}
where $\alpha(\omega)$ and $\mb \beta(\omega) = \{\beta_1(\omega), \cdots, \beta_P(\omega)\}^\intercal$ are effect functions, $\xi_j(\omega)$ are replicate-specific error functions such that $E\{\xi_j(\omega)\} = 0$ for all $\omega \in \mathbb{R}$ and $E\{\xi_j(\omega) \xi_j(\omega')\}=\sigma^2_\omega$ if $\omega=\omega'$ and 0 otherwise. Thus, we have the population-mean log-spectrum $E\{g_j(\mb X_j, \omega)\} = \alpha(\omega) + \mb X_j^\intercal \mb \beta(\omega)$.  Model \eqref{semilog} indicates that the replicate-specific log-spectra are random curves varying about a deterministic population-mean log-spectrum, given that these subjects possess the same fixed covariate. 	 It should be noted that \cite{krafty2011} considered a semiparametric mixed-effects model that can suffer from identifiability issues in certain scenarios. For example, their model is not identifiable when {\it deterministic} covariates are used to define a model of groups of independent time series with extra spectral variability. In contrast, our model is a fixed-effects model with stochastic covariates and does not have the identifiability issues present in the model of \cite{krafty2011}.

To estimate the effect functions, \cite{diggle1997} used a maximum likelihood approach, which involves numerically evaluating integrals of multivariate functions. \cite{iannaccone2001semiparametric} considered a state-space model formulation and used MCMC algorithms for estimations.  \cite{krafty2011} proposed a penalized least square approach with the smoothing parameters selected via cross-validations.   These approaches are computationally intensive, and thus may not be suitable for large datasets. In Sections \ref{sec:cep} and \ref{sec:lm}, we introduce a reparameterization of the model using cepstral coefficients representation of the log-spectra. This reparametrization leads to a cepstral multivariate linear model, enabling fast and accurate spectral analysis of replicated time series. This approach overcomes the computational challenges associated with the previous methods and is particularly well-suited for large datasets.

\subsection{Replicate-specific Cepstral Representation}
\label{sec:cep}

A natural representation of the log-spectrum is through the cepstrum. We approximate the replicate-specific log-spectra using the truncated cepstral coefficients, which provide a complete yet parsimonious characterization of the second-order frequency domain properties of the time series. The original definition of the cepstral coefficients was introduced by \cite{bogert1963} using the complex trigonometric polynomials. Since we consider real-valued time series and log-spectra that are periodic even functions, we define the cepstrum through a cosine series, as described in \cite{bloomfield1973}.

Define $\mathbb{G}$ as the space of even functions with period 1 whose first derivatives are square-integrable. When $g_j \in \mathbb{G}$ with probability 1, each replicate-specific log-spectrum possesses the cosine expansion,
\begin{eqnarray}\label{cepstral}
g_j(\mb X_j, \omega) = Y_{j0} + \sqrt{2}\sum_{k=1}^{\infty} Y_{jk} \cos(2 \pi \omega k), 
\end{eqnarray}	
where $(Y_{jk}: k=0,1,\cdots)$ are the replicate-specific cepstral coefficients with $Y_{j0} = \int_{-1/2}^{1/2}  g_j(\mb X_j, \omega) d\omega$ and $Y_{jk} = \int_{-1/2}^{1/2} g_j(\mb X_j, \omega) \sqrt{2}\cos(2 \pi \omega k) d \omega$.

The following regularity conditions are imposed on the cepstral coefficients
\begin{assumption}\label{ass1}
	$\sum_{k=1}^\infty k^2 \vert E(Y_{jk}) \vert^2 < \infty$ and $P\left (\sum_{k=1}^{\infty} k^2 |Y_{jk}|^2 < \infty\right ) = 1$ for $j=1, \cdots, N$.
\end{assumption}
Assumption \ref{ass1} implies that the mean replicate-specific log-spectra and error functions are absolutely continuous with square integrable first derivatives, and thus, error incurred by using only a finite number of cepstral coefficients is asymptotically negligible. This further indicates that only a finite number of cepstral coefficients are required  to represent the log-spectra \citep{bloomfield1973}.  As a result, the infinite-dimensional cepstral representation in Equation \eqref{cepstral} can be approximated by a finite-dimensional formulation for the log-spectrum by truncating the cosine series at some $K<T$,
\begin{equation*}\label{cepstral2}
g_j(\mb X_j, \omega) \approx Y_{j0} + \sqrt{2} \sum_{k=1}^{K-1} Y_{jk} \cos(2 \pi \omega k).
\end{equation*}	
Under this approximation, we denote the vector of truncated cepstral coefficients as $K$-vector $\mb Y_j= (Y_{j0}, \cdots, Y_{jK-1})^\intercal$.  In practice, we wish to choose a relatively small truncation number $K$ to save computational time and to present the frequency patterns of time series parismously without sacrificing accuracy. Our framework explores the spectral properties of time series using the truncated cepstral coefficients and utilize a cepstral-based multivariate linear model to establish a connection between the cepstral coefficients and covariates.


\subsection{Cepstral-based Multivariate Linear Model}
\label{sec:lm}
We impose the same smoothness properties on the effect functions and the replicate-specific log-spectra to ensure coherence in interpretation. Therefore, we model the effect functions $\alpha(\omega)$ and $\mb \beta(\omega) = \{\beta_1(\omega), \cdots, \beta_P(\omega)\}^\intercal$ in Equation \eqref{semilog} using the same cosine basis functions as those for the log-spectra. 
Define 
$\mb \phi(\omega) = \{ 1, \sqrt{2} \cos(2\pi \omega ), \cdots,  \sqrt{2}\cos[2\pi \omega (K-1)] \}^\intercal,$
the effect functions $\alpha(\omega)$ and $\beta_p(\omega)$ for $p=1,\dots,P$, and the error functions $\xi_j(\omega)$ can be expressed as 
\begin{equation*}\label{cepeffect}
\alpha(\omega) = \mb \phi(\omega)^\intercal \mathbf{A}, ~ \beta_p(\omega) = \mb \phi(\omega)^\intercal \mathbf{B}_p, ~\xi_j(\omega) = \mb \phi(\omega)^\intercal \mb \epsilon_j
\end{equation*}
respectively, where $\mathbf{A}$ and $\mathbf{B}_p$ are $K$-dimensional unknown coefficients, and $\mb \epsilon_j = (\epsilon_{j1}, \dots, \epsilon_{jK})^\intercal$. Further, the replicate-specific log-spectra can be expressed as $g_j(\mb X_j, \omega) = \mb \phi^\intercal(\omega) \mb Y_j$ and the right hand side of \eqref{semilog} is $\mb \phi(\omega)^\intercal \mathbf{A} + \sum_{p=1}^P X_{jp}\mb \phi(\omega)^\intercal \mathbf{B}_p + \mb \phi(\omega)^\intercal \mb \epsilon_j$.
Therefore, the semiparametric log-spectral model discussed \eqref{semilog} boils down to a cepstral-based multivariate linear regression model. Specifically, we have the following multivariate linear model that connects the truncated cepstral coefficients $\mb Y_j \in \mathbb{R}^K$ and covariates $\mb X_j\in \mathbb{R}^P$:
\begin{equation}\label{mlr}
\mb Y_j = \mathbf{A} + \mathbf{B}^\intercal \mb X_j + \mb \epsilon_j,
\end{equation}
where  $\mb \epsilon_j \in \mathbb{R}^K$ is mean zero and has covariance matrix $\mb \Sigma_{\epsilon}$, $\mathbf{A}\in\mathbb{R}^K$ is an unknown vector of intercept, and  $\mathbf{B} \in \mathbb{R}^{P\times K}$ is an unknown matrix of coefficients whose rows are $\mathbf{B}_1, \dots, \mathbf{B}_P$. Since $\mb Y_j$ approximately represents the low spectra $g_j(\mb X_j;\omega)$, each $\mathbf{B}_p$ quantifies the association between the $p$th covariate $X_{jp}$ and the second-order frequency patterns of the time series. More specifically, each element in $\mathbf{B}_p$ measures the relationship between $X_{jp}$ and specific frequency characteristics represented by the basis function $\mb \phi(\omega)$.

An important aspect of the proposed cepstral model is that it enables us to leverage the vast array of techniques available in the multivariate linear model literature to enhance the estimation and interpretation of the cepstral-based multivariate linear model for different applications.
For example, we can adopt existing sparse estimation procedures for the multivariate linear model, such as those in  \cite{lutz2006boosting,rothman2010sparse,wang2015joint}, if we believe the coefficient matrix $\mathbf{B}$ is sparse.  Alternatively, the reduced-rank regression and its extensions \citep[among others]{izenman1975reduced, chen2012reduced, chen2012sparse} can be used, if it is believed that the coefficient matrix $\mathbf{B}$ has a low-rank structure. Additionally, we can consider envelope regression models, which offer efficient estimation and improved predictive accuracy \citep{cook2010, su2011, cook2013} over the ordinary least squares. Last but not least, we can perform dimension reduction using sufficient dimension reduction approaches \citep{dong2022selective}.

\section{Estimation}\label{sec:est}

We develop a two-stage procedure for estimating the proposed model. In the first step, we estimate the truncated replicate-specific cepstral coefficients using a Whittle likelihood-based approach. In the second stage, we estimate the parameters of the cepstral-based multivariate linear model. We demonstrate the proposed estimation procedure by considering three estimators of the multivariate linear model, including the ordinary least squares estimator, the envelope estimator, and the reduced-rank estimator. It should be noted that other estimation procedures of the multivariate linear model can be easily adopted as well. Compared to other related methods, our estimation procedures are much faster as it does not require selecting smoothing parameters as in \cite{krafty2011} or involve complex MCMC algorithms as in \cite{wang2023adaptive}.

\subsection{Stage 1: Estimating the Cepstral Coefficients}\label{step1}

\RestyleAlgo{ruled}
\begin{algorithm}[h]
	\label{algo1}
	\caption{Fisher Scoring Algorithm for Estimating Cepstral Coefficients}\label{alg:two}
	{\bf Initialization}: Let $\mb L_j = [\log I_{j1} + \gamma, \cdots, \log(I_{jL}) + \gamma ]^\intercal$, where $\gamma \approx 0.577$ is the Euler-Mascheroni constant. Let $\mb \Phi$ be an $L \times K$ matrix with $\mb \phi_{\ell}$ as the $\ell$th row. For fixed $K$, calculate initial estimates of cepstral coefficients such that 
	$$\mb Y^{(0)}_j = (\mb \Phi^\intercal \mb \Phi)^{-1} \mb \Phi^\intercal \mb L_j.$$
	
	Let $\kappa$ be a small preselected threshold.
	
	\For{$j =1,\cdots, N$}{
		\While{${\cal L}_{jK} \left (\mb Y^{(m)}_j \right ) < \kappa$}{
			$$
			\mb Y^{(m+1)}_j = \mb Y^{(m)}_j +  \mathcal{U}^{-1}( \mb Y^{(m)}_j) \mathcal{I} (\mb Y^{(m)}_j),
			$$
			where
			$
			\mathcal{U} (\mb Y^{(m)}_j)= \sum_{\ell=1}^L \left [1- I_{j\ell} \exp(-\mb \phi^\intercal_\ell \mb Y^{(m)}_j) \right ] \mb \phi_{\ell},~~  \mathcal{I}( \mb Y^{(m)}_j) = \sum_{\ell=1}^L \mb \phi_{\ell}\mb \phi_{\ell}^\intercal
			$
			are score function and Fisher information matrix, respectively. 
		}
		Denote the final estimated replicate-specific cepstral coefficents as $\widehat{\mb Y}_j$.
	}
\end{algorithm}

For each time series, the replicate-specific periodogram is computed as
\begin{equation*}
I_{j \ell} = T^{-1} \left | \sum_{t=1}^T Z_{jt} \exp(- 2 \pi i \omega_{\ell} t)  \right |^2, ~~ j=1, \cdots, N,
\end{equation*}
where $\omega_{\ell} = \ell/T$, $\ell=1,\cdots, L$, are Fourier frequencies, and $L=\left \lfloor (T-1)/2 \right \rfloor$, which is the greatest integer that is less than or equal to $(T-1)/2$. It is known that $I_{j \ell}$ are approximately independent and distributed as $\exp\{g_j(\mb X_j, \omega_{\ell})\} \chi_2^2/2$ when $T$ is sufficiently large \citep{brillinger2001}. This large sample distribution of the periodograms leads to the Whittle likelihood \citep{whittle1953} for the cepstral coefficients of the replicate-specific log-spectra. 
Thus, for fixed $K$, the estimates of the replicate-specific cepstral coefficents can be obtained by minimizing the negative log-Whittle likelihood
\begin{eqnarray*}
{\cal L}_{jK} (\mb Y_j) =  \sum_{\ell=1}^L \left \{  I_{j\ell} \exp(-\mb \phi_\ell^\intercal \mb Y_j)  +  \mb \phi_\ell^\intercal \mb Y_j\right  \},
\end{eqnarray*}
where $\mb Y_j = (Y_{j0}, \cdots, Y_{jK-1})^\intercal$ and $\mb \phi_\ell = \mb \phi(\omega_\ell)$.
The estimated replicate-specific cepstral coefficients $\widehat{\mb Y}_j$ for $j=1,\cdots,N$ are obtained using a Fisher scoring algorithm outlined in Algorithm \ref{algo1}. 	Note that, according to Theorem 1 in \cite{krafty2011}, the Euler-Mascherroni constant $\gamma \approx 0.5777$ is needed when computing the initial bias-corrected log periodograms for obtaining accurate estimates of the cepstral coefficients.

%
 In practice, the truncating number $K$ is selected through AIC by minimizing
$$
AIC(K) = \sum_{j=1}^{N} {\cal L}_{jK} (\widehat{\mb Y}_j) + 2NK,
$$
where $	{\cal L}_{jK} (\widehat{\mb Y}_j) = \sum_{\ell=1}^L \left \{  I_{j\ell} \exp(-\mb \phi^\intercal_\ell \widehat{\mb Y}_j)  +  \mb \phi^\intercal_\ell \widehat{\mb Y}_j\right  \}$ and $\widehat{\mb Y}_j = (\widehat{Y}_{j0}, \cdots, \widehat{Y}_{jk-1})^\intercal$ are an vector of the estimates of cepstral coefficients. Other ways of selecting $K$, such as the generalized cross-validation approach of \cite{ombao2001simple}, can also be adopted.  However, we use the AIC-based approach due to its good performance in our simulations and its low computational cost. Proposition \ref{cepconsistent} establishes the asymptotic properties of the  estimates of the replicate-specific cepstral coefficients 
\begin{proposition}\label{cepconsistent}
 Under Assumptions \ref{assum_spectral} and \ref{ass1}, the Whittle likelihood-based estimates of the replicate-specific cepstral coefficients are consistent such that $\widehat{\mb Y}_j \overset{p}{\to} \mb Y_j$ and $\sqrt{T} (\widehat{\mb Y}_j - \mb Y_j)\overset{d}{\to} N(0,\mathbf{V})$, where $\mathbf{V}^{-1}=\int_{-1/2}^{1/2} \mb \phi_\ell \mb \phi_\ell^\intercal$.
\end{proposition}	

\subsection{Stage 2: Estimating Parameters of Multivariate Linear Model}\label{step2}

We first express the cepstral-based multivariate linear model in matrix form
$
\mathbf{Y}= \mathbf{A} + \mathbf{X} \mathbf{B} + \mb \epsilon, 
$
where $\mathbf{Y}$ is the $N \times K$ cepstral coefficients matrix, $\mathbf{X}$ is the $N \times P$ matrix of covariates, $\mathbf{B}$ is the $P \times K$ matrix of regression coefficients, $\mb \epsilon$ is the $N \times K$ error matrix.
Once the estimates of replicate-specific cepstral coefficients $\widehat{\mathbf{Y}}$ are obtained, we compute the estimates of the parameters of the multivariate linear model, namely, $\widehat{\mathbf{A}}$ and $\widehat{\mathbf{B}}$,  which allows us to obtain the estimates of effect functions  $\widehat{\mb \alpha}(\omega)$ and  $\widehat{\mb \beta}(\omega)$. At this stage, various approaches for estimating the multivariate linear model can be applied, depending on the problem of interest or applications. In this section, we discuss three commonly used estimators, including the ordinary least square estimator, the reduced-rank regression \citep{izenman1975reduced}, and the envelope estimator \citep{cook2013}.


\subsubsection{The Ordinary Least Square Estimator}

The ordinary least squares estimator (OLS) of model \eqref{mlr} is
$
\widehat{\mathbf{B}}_{ols}  = (\mathbf{X}^\intercal\mathbf{X})^{-1} \mathbf{X}^\intercal \widehat{\mathbf{Y}},
$
where  $\widehat{\mathbf{Y}}$ is the $N \times K$ estimated cepstral coefficients matrix obtained by the minimizing the negative log-Whittle likelihood as in Section \ref{step1}. The OLS-based estimator of the effect functions are then given by
$$\widehat{\mb \beta}_{ols}(\omega) = \{\widehat{\beta}_{ols,P}(\omega), \cdots, \widehat{\beta}_{ols,P}(\omega)\}^\intercal~\text{with}~ \widehat{\beta}_{ols,p}(\omega) = \mb \phi(\omega)^\intercal \widehat{\mathbf{B}}_{ols,p}, ~p=1, \cdots, P$$
where $\widehat{\mathbf{B}}_{ols,p}$ is the $p$th row of $\widehat{\mathbf{B}}_{ols}$. The following additional assumptions are made for the pointwise consistency of the proposed OLS-based two-stage estimator. 
\begin{assumption}\label{ass2}
	The fourth moment of $\mathbf{X}$ exists and there is a positive definite matrix $\mathbf{Q}$ such that $\mathbf{X}\mathbf{X}^\intercal/N \rightarrow \mathbf{Q}$ as $N \rightarrow \infty$.
\end{assumption}	
\begin{theorem}\label{thm:ols}
With Assumptions \ref{assum_spectral}, \ref{ass1} and \ref{ass2}, as $N,T \rightarrow \infty$, we have
	$$
	\sup_{p=1,\cdots, P; ~\omega \in \mathbb{R}} E\left \{\left |\widehat{\beta}_{ols,p}(\omega)  - \beta_p(\omega) \right |^2 \right \} = O(T^{-1}) + O(N^{-1}).
	$$
\end{theorem}	 
The estimates $\widehat{\mathbf{B}}_{ols}$ can be also obtained by performing $K$ separate OLS regressions, where the regression coefficients from the $k$th regression form the $k$th column of $\widehat{\mathbf{B}}_{ols}$. However, it is important to note that the OLS estimator ignores the correlation among the responses, which may lead to suboptimal results in certain situations.





\subsubsection{The Reduce-Rank Regression}

One way to account for possible interrelationships between the responses is imposing a rank constraint on $\mathbf{B}$. The reduced-rank regression model of \cite{izenman1975reduced} considers
\begin{equation*}
\mathbf{B} = \mathbf{C} \mathbf{D}^\intercal,   ~\text{rank}(\mathbf{B}) =m, ~\text{and}~ m < \min(P,K),
\end{equation*}
where  $\mathbf{C} \in \mathbb{R}^{P\times m}$ and  $\mathbf{D} \in \mathbb{R}^{K \times m}$. Note that $\mathbf{X} \mathbf{C}$ is of reduced dimension with only $m$ components. These $m$ linear combinations of the covariates can be interpreted as unobservable latent factors that drive the variability in the cepstral coefficients. It is easy to see that 
$
\sps(\mathbf{B}_{rrr}) \subset \sps( \mb \Sigma_{XX}^{-1}\mb \Sigma_{XY}) = \sps(\mathbf{B}_{ols}),
$
where $\sps(\mathbf{B})$ denotes the column space of $\mathbf{B}$. For fixed $m$, the rank-constrained $\mathbf{B}$ is estimated by
\begin{equation*}\label{eqn:rrrr}
 \min_{\mathbf{C}, \mathbf{D};\text{rank}(\mathbf{C})=\text{rank}(\mathbf{D})=m}  || \mathbf{Y} - \mathbf{X} \mathbf{C} \mathbf{D}^\intercal ||^2.
\end{equation*}
From Theorem 1 of \cite{izenman1975reduced}, the solution of this minimization problem leads to the reduced-rank-based estimator
$$
\widehat{\mathbf{B}}_{rrr} =  \widehat{\mathbf{C}} \widehat{\mathbf{D}}^\intercal = (\mathbf{X}^\intercal\mathbf{X})^{-1} \mathbf{X}^\intercal \widehat{\mathbf{Y}} \left ( \sum_{j=1}^m \hat{\mb u}_j \hat{\mb u}_j^\intercal \right),
$$
where  $\widehat{\mathbf{Y}}$ is the $N \times K$ estimated cepstral coefficients matrix obtained by the minimizing the negative log-Whittle likelihood as in Section \ref{step1} and $ \hat{\mb u}_j $ is the eigenvector corresponding to the $j$th leading eigenvalue of $\widehat{\mb \Sigma}_{X\widehat{Y}}^\intercal  \widehat{\mb \Sigma}_{XX}^{-1} \widehat{\mb \Sigma}_{X\widehat{Y}}$, wehre $\widehat{\mb \Sigma}_{XX}$ is the sample covariance matrix of $\mathbf{X}$ and $\widehat{\mb \Sigma}_{X\widehat{Y}}$ is the sample covariance between $\mb X$ and $\widehat{\mb Y}$. Finally, the reduced-rank-based estimator of the effect functions are 
$$\widehat{\mb \beta}_{rrr}(\omega) = \{\widehat{\beta}_{rrr,P}(\omega), \cdots, \widehat{\beta}_{rrr,P}(\omega)\}^\intercal~\text{with}~ \widehat{\beta}_{rrr,p}(\omega) = \mb \phi(\omega)^\intercal \widehat{\mathbf{B}}_{rrr,p}, ~p=1, \cdots, P,$$
where $\widehat{\mathbf{B}}_{rrr,p}$ is the $p$th row of $\widehat{\mathbf{B}}_{rrr}$. 
The pointwise consistency of the reduced-rank-based estimator is eastablised 
\begin{proposition}
	With Assumptions \ref{assum_spectral}, \ref{ass1} and \ref{ass2}, and fixed $m$, as $N, T \rightarrow \infty$, we have
	$$
	\sup_{p=1,\cdots, P; ~\omega \in \mathbb{R}} E\{|\widehat{\beta}_{rrr,p}(\omega)  - \beta_p(\omega) |^2 \} = O(T^{-1}) + O(N^{-1}).
	$$
\end{proposition}

\subsubsection{The Envelope Estimator}
The original envelope model was first developed by \cite{cook2010} for parsimonious parameterizations of responses in the multivariate linear model. However, we are more concerned with reducing covariates dimension. Thus, we consider the predictor envelope model of \cite{cook2013}. Denote $\mathcal{R}$ as a subspace of  $\mathbb{R}^P$ and let $\mathcal{R}^\perp$ be its orthogonal complement in $\mathbb{R}^P$.  The predictor envelope regression model is 
\begin{equation}\label{enve}
\mathbf{Y}= \mathbf{X} \mathbf{B} + \mb \epsilon, ~\text{with}~\sps(\mathbf{B}) \in \mathcal{R}, ~\mb \Sigma_{XX} \mathcal{R} \in \mathcal{R}, ~\text{and} ~\mb \Sigma_{XX} \mathcal{R}^\perp \in \mathcal{R}^\perp.
\end{equation}
Model \eqref{enve} implies that  $\mathcal{R}$ is a reducing subspace of $\mb \Sigma_{XX}$ that contains $\sps(\mathbf{B})$. The intersection
of all such reducing subspaces is known as the $\mb \Sigma_{XX}$-envelope of $\sps(\mathbf{B})$, and is denoted by $\xi_{\mb \Sigma_{XX}} \{\mathcal{B}\}$. Let $r = \dim\{\xi_{\mb \Sigma_{XX}} \{\mathcal{B}\}\}$ and $(\mb \Gamma, \mb \Gamma_0) \in \mathbb{R}^{P \times P}$.  Suppose  $\mb \Gamma \in \mathbb{R}^{P\times r}$ with $r \le P$ is a semi-orthogonal basis matrix for $\xi_{\mb \Sigma_{XX}} \{\mathcal{B}\}$ 
and $\mb \Gamma_0 \in \mathbb{R}^{P \times (P-r)}$. The envelope model assumes that $\mb \Gamma$ and $\mb \Gamma_0$ satisfy the conditions  $\cor(\mb \Gamma^\intercal \mb X, \mb \Gamma_0^\intercal\mb X) = 0$ and $\cor(\mb Y, \mb \Gamma_0^\intercal \mb X|\mb \Gamma^\intercal \mb X) = 0$. Consequently, $
\mb \Sigma_{XX} = \mb \Gamma \mb \Delta \mb \Gamma^\intercal + \mb \Gamma_0 \mb \Delta_0 \mb \Gamma_0^\intercal, 
$
where $\mb \Delta = \mb \Gamma^\intercal \mb \Sigma_{XX} \mb \Gamma \in \mathbb{R}^{r \times r}$ and $\mb \Delta_0 = \mb \Gamma_0^\intercal \mb \Sigma_{XX} \mb \Gamma_0 \in \mathbb{R}^{(P-r) \times (P-r)}$. This means that $\mb \Gamma^\intercal \mb X$ is the material part that contains all the information that is available about $\mathbf{B}$ from $\mb X$, whereas $\mb \Gamma_0^\intercal \mb X$ is the immaterial part of $\mb X$. Various extensions of Model \eqref{enve} have been developed, we refer to \cite{cook2018introduction} for a comprehensive review. The envelope-based estimator of $\mathbf{B}$ is given by
\begin{equation*}
\widehat{\mathbf{B}}_{env} = \widehat{\mb \Gamma} \left[\widehat{\mb \Gamma}^\intercal  (\mathbf{X}^\intercal \mathbf{X})^{-1} \widehat{\mb \Gamma} \right ]^{-1} \widehat{\mb \Gamma} \mathbf{X}^\intercal \widehat{\mathbf{Y}} , 
\end{equation*}
where the estimates $\widehat{\mb \Gamma}$ can be obtained by minimizing the log likelihood under the orthogonality constraints \citep{wen2013feasible}. We refer readers to \cite{cook2013} for more details. Then, the envelope-based estimator of the effect functions are
$$\widehat{\mb \beta}_{env}(\omega) = \{\widehat{\beta}_{env,P}(\omega), \cdots, \widehat{\beta}_{env,P}(\omega)\}^\intercal~\text{with}~ \widehat{\beta}_{env,p}(\omega) = \mb \phi(\omega)^\intercal \widehat{\mathbf{B}}_{env,p}, ~p=1, \cdots, P$$
where $\widehat{\mathbf{B}}_{env,p}$ is the $p$th row of $\widehat{\mathbf{B}}_{env}$. 
The envelope estimation procedure can be implemented through the \texttt{R} package \texttt{Renvlp}.  The following proposition provides the pointwise consistency of the proposed envelope-based estimator. 
\begin{proposition}\label{thm:env}
	With Assumptions \ref{assum_spectral}, \ref{ass1} and \ref{ass2}, and fixed $r$, as  $N,T \rightarrow \infty$, we have
	$$
	\sup_{p=1,\cdots, P; ~\omega \in \mathbb{R}} E\{|\widehat{\beta}_{env,p}(\omega)  - \beta_p(\omega) |^2 \} = O(T^{-1}) + O(N^{-1}).
	$$
\end{proposition}

There are several situations in which the envelope-based estimator might be superior to the OLS-based estimator as outlined in \cite{cook2013}.  First, the variance of $\vect(\widehat{\mathbf{B}}_{ols})$ could be much larger than that of $\vect(\widehat{\mathbf{B}}_{env})$ especially when  $r$ is small relative to $P$, reflecting the gain in estimation accuracy through dimension reduction. Second, the variance of $\vect(\widehat{\mathbf{B}}_{ols})$ could be much larger than that of $\vect(\widehat{\mathbf{B}}_{env})$ if the collinearity among $\mathbf{X}$ is strong and is associated with the immaterial variation. Lastly, the envelope-based estimator is preferred when the signal-to-noise ratio is small.

\subsection{Bootstrap Confidence Interval for Effect Functions}
Following \cite{krafty2011}, we construct pointwise confidence intervals of the effect functions using a residual-based bootstrap procedure which is commonly used in regression models \citep{davison1997bootstrap}.
 First, we generate the $b$th bootstrap error terms, $\hat{\xi}_j^{(b)}(\omega) = \mb \phi(\omega)^\intercal \hat{\mb \epsilon}_j^{(b)}$ by randomly sampling from the estimated residuals $\hat{\xi}_j(\omega) = \mb \phi(\omega)^\intercal \hat{\mb \epsilon}_j$ for $j=1,\cdots,N$, where $\hat{\mb \epsilon}_j$ is estimated residuals of the multivariate linear regression. Second, according to Theorem 2 in \cite{dai2004},  we generate the $b$th bootstrap sample of the collection of time series by simulating time series epochs $Z_{jt}^{(b)}$, for $j=1, \cdots, N$ and $t=1, \cdots, T$, given the estimated replicate-specific spectra $$\hat{f}_j^{(b)}(\mb X_j, \omega)= \exp[\hat{g}_j^{(b)}(\mb X_j, \omega)]= \exp\{ \widehat{\alpha}(\omega) + \mb X_j^\intercal \widehat{\mb \beta}(\omega) + \hat{\xi}_j^{(b)}(\omega)\}.$$
Then, we obtain the $b$th estimates of effect functions based on the $b$th bootstrap sample using the proposed two-stage estimators introduced in Section 3, and denote the estimates as $\widehat{\mb \beta}^{(b)}(\omega) =[\widehat{\beta}_1^{(b)}(\omega) , \cdots \widehat{\beta}_P^{(b)}(\omega)  ]^\intercal$. Finally, we estimate the $(1-\alpha)\%$ confidence interval for $\beta_p(\omega)$, $p =1,\cdots, P$, as $[\zeta_p(\omega; \alpha/2), \zeta_p(\omega; 1-\alpha/2)],$ where $\zeta _p(\omega; \alpha/2)$ is the $\alpha$ percentile of the set $\{ \widehat{\beta}_p^{(b)}(\omega) - \text{Bias}_p (\omega)\}_{b=1}^B$, and $\text{Bias}_p (\omega)= B^{-1}\sum_{b=1}^B \widehat{\beta}_p^{(b)}(\omega)- \widehat{\beta}_p(\omega)$.

\section{Simulation Studies}


Simulation studies are conducted to investigate the empirical performance of the proposed model and compare it to the performance of two existing methods. The first existing method is the functional semiparametric model of \cite{krafty2011}. This approach estimates the model by using the penalized least squares, and we denote it as PLS. It is important to note that PLS allows for modeling of mixed effects, which is a more general setting than what is considered in this work. Accordingly, we implement a simplified version with only fixed effects included. The second existing method is the Bayesian sum of trees model of \cite{wang2023adaptive}, which we denote as Tree. The proposed model was estimated by the two-stage estimation procedure introduced in Section \ref{sec:est}, where the ordinary least squares, envelope, and reduced-rank estimators are used in the second stage. We denote them as cepOLS, cepENV, and cepRRR, respectively. For all three proposed methods, the truncation numbers are estimated by the AIC procedure discussed in Section \ref{step1}. Performance of the estimation procedures is evaluated through the square error of the log-spectral effect functions averaged over the Fourier frequencies (ASE). For example, the ASE of $\widehat{\beta}_1$ is given by
$
L^{-1}\sum_{\ell=1}^L   [ \widehat{\beta}_1(\omega_\ell) - \beta_1(\omega_\ell)  ]^2.
$ 
Since the method of \cite{wang2023adaptive} is nonparametric in nature and is not able to produce effect functions, we additionally consider the ASE of fitted replicate-specific log-spectra,
$(NL)^{-1}\sum_{\ell=1}^L \sum_{j=1}^N [ \hat{g}(\mb X_j, \omega_{\ell}) - g(\mb X_j, \omega_{\ell}) ]^2$. 


\subsection{Simulation Settings}
\subsubsection{Example 1}

This example is adopted from the simulation study of \cite{krafty2011} with a slight modification that excludes the random effects. In particular, we set the replicate-specific log-spectra to be $g_j(\mb X_j, \omega) = \alpha(\omega) + X_{j1} \beta_1(\omega) + \xi_j (\omega)$ for $\alpha(\omega)= 2\cos(2\pi\omega)$, $\beta_1(\omega)= 2\cos(4\pi\omega)$, and $\xi_j (\omega)= \epsilon_{j1} + \epsilon_{j2}\cos(2\pi  \omega) + \epsilon_{j3} \cos(4 \pi  \omega)$, where $X_{j1}$ are independent uniform random variables over $[0,1]$ and $(X_{j2}, \cdots, X_{jP})$ are independent multivariate normal random variables with zero-mean and unit variance, and $\epsilon_{j1}$, $\epsilon_{j2}$, $\epsilon_{j3}$ are generated from zero-mean Normal distribution with variance 0.5.   After a replicate-specific log-spectrum is simulated, the square root of the spectrum is calculated and used as the replicate-specific transfer function to simulate the time series $Z_{jt}$ according to \cite{dai2004}. 
Random samples of $N$ independent units of time series epochs of length $T$ are drawn for the eight possible combinations of $P=1, 10$, $N = 50, 100$, and $T = 50, 100$. Note that $(X_{j2}, \cdots, X_{jP})$ are noise variables when $P=10$ and there is no noise variable when $P=1$. For the proposed methods, only cepOLS are used when $P=1$ since cepENV and cepRRR are only applicable to multiple covariates.

\subsubsection{Example 2}

We consider a more complex example where the replicated-specific log-spectra are $g_j(\mb X_j, \omega) = \alpha(\omega) + X_{j1} \beta_1(\omega) + X_{j2} \beta_2(\omega)  + \xi_j (\omega)$ for $\alpha(\omega)= 2\cos(2\pi\omega)$, $\beta_1(\omega)= 2\cos(4\pi\omega) + 2\cos(6\pi\omega)$, $\beta_2(\omega)= 2\cos(8\pi\omega)$, and $\xi_j (\omega)= \epsilon_{j1} + \epsilon_{j2}\cos(2\pi  \omega) + \epsilon_{j3} \cos(4 \pi  \omega)$, where $\mb X_j$ are independent $P$-dimension multivariate normal random variables with mean zero and covariance matrix $\mb \Sigma_{XX}$ with elements $\sigma_{pq} = \tau^{|p-q|}$ for $1 \le p,q \le P$, $P=10$, and $\epsilon_{j1}$, $\epsilon_{j2}$, $\epsilon_{j3}$ are generated from zero-mean Normal distribution with variance 0.5. Based on the replicate-specific log-spectra, time series $Z_{jt}$ are then simulated \citep{dai2004}. 
Random samples of $N$ independent units of time series epochs of length $T$ are drawn for the eight possible combinations of $\tau=0, 0.5$, $N = 50, 100$ and $T = 50, 100$.



\subsection{Simulation Results}

All simulations were carried out on a Windows 10 machine equipped with a 3.6 GHz Intel Core i7 processor and 32 GB RAM. All results in this section are based on 500 repetitions. Table \ref{tab:sim1} and Table \ref{tab:sim2} display the means and standard deviations of ASEs for estimating $\alpha(\omega)$, $\mb \beta(\omega)$, and log-spectra $g(\mathbf{X}_j, \omega)$, along with the mean computational times. The average across the curve coverage of the 95\% pointwise bootstrap confidence intervals for the two examples across different settings ranged from 93.4\% to 97.5\%. Additional simulation results on the accuracy of the proposed AIC-based approach for selecting the truncation number $K$ are provided in the Supplementary Materials.

  For Example 1, several important observations can be made. First, in each of the eight settings, the proposed methods, cepOLS, cepENV, and cepRRR-have smaller ASEs than the other methods for estimating effect functions and log-spectra. Second, when $P=10$, all three proposed methods show similar performance in estimating effect functions. However, cepRRR and cepENV exhibit smaller ASEs when estimating log-spectra.  This difference in performance is likely due to the noise variable filtering and dimension reduction capabilities inherent in the reduced rank regression and the envelope model.  Third, the proposed methods are significantly faster than the other two methods, especially when $N$, $T$, or $P$ is large. It's worth noting that the proposed methods only experience a slight increase in computation time with increases in $N$, $T$, and $P$, indicating their scalability to large datasets, while the other methods become computationally infeasible.

\begin{table}[h]
	\caption{Simulation results of Example 1. Square error $\times 10^2$ averaged over the Fourier frequencies are displayed for $\alpha$, $\beta$, and log-spectra. The average computational time in minutes are also reported.}
	\label{tab:sim1}
	\vspace{0.3cm}
	\centering
	\scalebox{0.7}{
		\begin{tabular}{ccccccccccccc}
			\hline
			&    & ~  &  \multicolumn{3}{c}{$P=1$} & ~& \multicolumn{5}{c}{$P=10$} \\
			\cline{4-6} \cline{8-12}
			$T$	& $N$ & & cepOLS & PLS & Tree & & cepOLS& cepENV& cepRRR& PLS & Tree  \\  
			\hline
			&  & $\widehat{\alpha}(\omega)$ & 6.01 (5.09)    &  10.62 (6.36)  & ---  & ~&6.72 (5.88)   &  7.44 (6.46)   & 7.60 (7.42)  &  10.80 (6.56)  & --- \\
			50 & 50 & $\widehat{\beta}_1(\omega)$ & 16.12 (12.58)   &  20.50 (15.12)   & --- & & 19.14 (17.83)   &  21.26 (21.28)   & 22.15 (20.59)  &  25.52 (16.32) & --- \\
			&  & $\widehat{g}(\mb X_j, \omega)$ &3.33 (2.11)  &  10.70 (5.01)  & 11.75 (8.28) & & 7.71 (3.80)   &  5.66 (3.01)  &7.37 (3.50)   & 11.39 (6.43)  & 16.32 (9.26) \\
			&  & Time &  0.01 &   3.73 & 31.70 &  &   0.01  &  0.22 & 0.02 &  8.54  & 39.51  \\
			\hline
			&  & $\widehat{\alpha}(\omega)$ & 3.05 (2.81)  &  10.28 (4.18)   & --- & & 3.14 (2.79)  &  3.32 (2.93)  & 3.22 (2.90)   &  10.30 (5.18)  & ---\\
			50 & 100 & $\widehat{\beta}_1(\omega)$ & 7.57 (6.11)    &  19.41 (14.55) & ---& & 8.55 (6.39)   & 9.10  (6.75)   & 8.99 (7.75)   &  20.49 (15.43)  & --- \\
			&  & $\widehat{g}(\mb X_j, \omega)$ & 1.80 (1.20)    & 8.56 (3.77)   & 9.18 (5.98) & & 3.49 (2.03)   &  2.90 (1.45)   & 3.56 (1.50)  &  9.93 (3.74) & 14.56 (6.97)\\
			&  & Time  & 0.02  &  4.04  & 57.90 & &  0.02  &  0.18 & 0.02 &  11.08 & 66.15 \\
			\hline
			&  & $\widehat{\alpha}(\omega)$& 4.93 (4.13)    & 6.85 (5.39)   & --- &  & 5.85 (5.32)   &  6.16 (5.44)   & 6.44 (6.12)  &  8.58 (7.31)  & ---\\
			100 & 50 & $\widehat{\beta}_1(\omega)$ & 13.27 (12.92)   & 21.01 (13.22)   & ---& & 16.33 (14.09)   &  17.31 (15.19)   & 18.60 (19.94)   &  21.41 (14.04)   & --- \\
			&  & $\widehat{g}(\mb X_j, \omega)$&2.48 (1.46)    &  10.02 (4.37) & 9.98 (7.97) &  & 6.69 (3.20)   &  4.72 (3.54)  & 5.99 (2.72)   & 10.85 (4.67)  &  15.19 (7.47)\\
			&  & Time  &  0.02 &  36.86  & 53.63 & & 0.02   & 0.20  & 0.02 &  82.46  & 60.25 \\			
			\hline			
			&  & $\widehat{\alpha}(\omega)$ & 2.52 (2.08)    &  3.67 (3.31)  & ---& & 2.86 (2.46)   &  3.06 (2.56)   & 3.01 (2.55)   &  4.92 (3.76)   & --- \\
			100 & 100 & $\widehat{\beta}_1(\omega)$& 6.83 (6.31)    &  16.83 (12.01)  & ---&  & 7.85 (6.83)   &  8.56 (7.43)  & 8.56 (7.48)   & 17.71 (13.43)   & --- \\
			&  & $\widehat{g}(\mb X_j, \omega)$& 1.32 (0.82)   &  6.83 (2.45)  & 8.83 (5.05) &  & 4.36 (1.70)  &  2.33 (1.12)   & 2.94 (1.19)  & 8.74 (3.01)  & 12.01 (6.01)\\
			&  & Time &  0.03 &  38.48   & 95.41 & & 0.03   & 0.17  & 0.03 &  98.07  & 98.93 \\			
			\hline		
		\end{tabular}
	}
\end{table}

In Example 2, the proposed methods outperform the other two methods in estimating effect functions and log-spectra across all settings. Note that the differences in ASEs between the proposed methods and the other two methods are bigger than those in Example 1, illustrating the advantages of using the proposed methods for more complex models. The tree-based model from \cite{wang2023adaptive} exhibits the largest ASEs and the longest computational time.  This is not surprising, given that the Tree method is nonparametric, which is less suitable for a complex semiparametric model. It's worth noting that the proposed cepRRR shows a large ASE when $N=50$ because it produces extremely large squared errors in some repetitions due to convergence issues. This suggests that cepRRR might require a larger sample size to produce stable estimation results. Moreover, cepOLS and cepENV yield similar ASEs when $\tau=0$. But, cepENV achieves the smallest ASEs when $\tau=0.5$. This result aligns with findings in \cite{cook2013}, which suggest that the envelope estimator's accuracy gains over the OLS estimator can be significant when collinearity among covariates is moderate or strong. Therefore, for datasets with a relatively small sample size and strong correlations, such as the COP trajectory dataset considered in Section \ref{sec:application}, we recommend using cepENV for stable and accurate estimation. Again, the proposed methods have significantly shorter computational time than the other two methods in all settings.

\begin{table}[h]
	\caption{Simulation results of Example 2. Square error $\times 10^2$ averaged over the Fourier frequencies are displayed for $\alpha$, $\beta$, and log-spectra. The average computational time in minutes are also reported.}
	\label{tab:sim2}
	\vspace{0.3cm}
	\centering
	\scalebox{0.63}{
		\begin{tabular}{cccccccccccccccc}
			\hline
&     ~  &  \multicolumn{5}{c}{$\tau=0$} & ~& \multicolumn{5}{c}{$\tau=0.5$} \\
			\cline{3-7} \cline{9-13}
$T, N$ & & cepOLS& cepENV& cepRRR& PLS & Tree & & cepOLS& cepENV& cepRRR& PLS & Tree  \\  
\hline
&   $\widehat{\alpha}(\omega)$ & 3.29 (2.03)   & 2.92 (1.81) & 3.16 (2.60) & 10.01 (4.61) & ---  & & 3.27 (2.05)  & 2.99 (1.85)   & 4.26 (4.33) &  10.11 (4.99)  & --- \\
 &   $\widehat{\beta}_1(\omega)$ & 1.99 (1.39)     & 1.82 (1.35) &  1.92 (1.62) & 6.29 (5.45) & ---  & ~& 2.91 (2.29) & 2.44 (1.84)   & 27.29 (47.05) & 6.48 (5.83)   & --- \\
50, 50 & $\widehat{\beta}_2(\omega)$ & 1.71 (1.28)    & 1.56 (1.18) &  5.49 (27.62) &  8.74 (5.76) & ---  & ~& 3.16 (2.28) &  2.80 (2.35)  &  33.92 (63.58))&  8.85 (6.11)  & --- \\
&   $\widehat{g}(\mb X_j, \omega)$ & 10.87 (3.28)     & 8.15 (2.65) & 17.57 (3.63) & 36.01 (6.88)  &  77.91 (25.82) & & 17.88 (3.77) &  8.69 (2.84)  & 47.09 (48.70) &  36.18 (7.51)  & 78.94 (25.91) \\
&   Time&  0.02    & 0.31 & 0.03& 10.90 & 24.25  & ~&   0.03   & 0.38 &  0.03 &  12.26 & 23.67   \\
\hline
&   $\widehat{\alpha}(\omega)$ & 2.18 (1.69)   & 2.09 (1.54) & 2.07 (1.53) & 8.98 (3.28) & ---  & ~&2.13 (1.62)  &  2.09 (1.53)  & 2.09 (1.52) &  9.98 (3.25)  & --- \\
&   $\widehat{\beta}_1(\omega)$ & 1.19 (1.20)     & 1.12 (1.12) & 1.10 (1.09) & 6.09 (4.98)  & ---  & ~& 1.50 (1.27) &  1.42 (1.15)  &1.43 (1.17)  &  6.02 (4.79)  & --- \\
50, 100 & $\widehat{\beta}_2(\omega)$ &0.74 (0.54)     & 0.69 (0.57) &  0.70 (0.56)&  8.21 (6.02)& ---  & ~& 1.48 (1.23) &  1.29 (1.05)  & 1.30 (1.08) &  8.16 (6.31)   & --- \\
&   $\widehat{g}(\mb X_j, \omega)$ & 5.08 (1.91)     & 5.04 (0.98) & 5.24 (1.25) &  35.17 (4.89) & 69.11 (26.24)  & ~& 10.32 (2.42) &  5.39  (2.10)  &  5.42 (2.03)& 35.69 (5.02) & 70.21 (26.92)\\
&   Time &  0.04    & 0.38 & 0.04& 14.41 &  34.94 & ~& 0.04 &  0.42  & 0.04  &  14.25  & 35.98   \\
\hline
&  $\widehat{\alpha}(\omega)$ & 2.18 (1.81)   & 1.78 (1.58) & 1.80 (1.56) & 3.39 (3.32) & ---  & ~&2.23 (1.74) &  1.82 (1.50)  & 3.21 (5.01) &  3.48 (4.31) & --- \\
&   $\widehat{\beta}_1(\omega)$ & 1.82 (1.46)    & 1.54 (1.18) & 1.58 (1.22) & 4.76 (3.21) & ---  & ~& 2.66 (2.12) &  2.22 (2.14)  & 26.85 (50.11)  & 5.26 (3.92)   & --- \\
100, 50 & $\widehat{\beta}_2(\omega)$ & 1.47 (1.09)     & 1.19 (0.92) &  3.22 (19.81) & 5.89 (3.77) & ---  & ~& 2.67 (2.42) & 1.84 (1.49)   & 28.01 (60.22) &  6.68 (4.02) & --- \\
&   $\widehat{g}(\mb X_j, \omega)$ & 7.46 (4.88)     & 5.32 (2.37) & 14.36 (3.50) & 29.86 (4.11) & 75.10 (18.51) & ~& 14.48 (3.33) &  5.92 (2.19)  & 45.10 (50.79) & 30.03 (4.56) & 76.10 (18.50)\\
&   Time &  0.04    & 0.39 & 0.03& 17.14 & 30.21  & ~& 0.04 &  0.43  &  0.04 &  18.22  & 29.37  \\
\hline
&   $\widehat{\alpha}(\omega)$ & 1.12 (1.02)   & 1.04 (0.91) & 1.04 (0.91) &  3.32 (2.25) & ---  & ~& 1.09 (0.97) &  1.02 (0.91)  & 1.03 (0.91) &  3.39 (2.43) & --- \\
&   $\widehat{\beta}_1(\omega)$ & 0.87 (0.68)     & 0.82 (0.61) & 0.82 (0.62) & 4.56 (3.12)& ---  & ~& 1.13 (0.80) &  1.10 (0.84)  &1.11 (0.86)  &  4.84 (3.41)  & --- \\
100, 100 & $\widehat{\beta}_2(\omega)$ & 0.83 (0.68)     & 0.74 (0.58) &  0.75 (0.60) & 5.29 (3.54)  & ---  & ~& 1.25 (1.14) & 0.91 (0.73)    &  0.94 (0.76)&  5.46 (3.65)  & --- \\
&   $\widehat{g}(\mb X_j, \omega)$ & 3.48 (1.32)     & 3.10 (1.30) & 3.21 (1.38) &  27.75 (3.96) &  58.37 (14.01) & ~& 7.62 (1.97) &  3.23 (1.36)  & 3.59 (1.39) & 28.38 (4.23) & 58.66 (14.00) \\
&   Time &  0.06   & 0.40 & 0.07& 18.65 & 40.54  & ~& 0.07  &  0.50  & 0.08 &  19.01  & 43.51  \\
\hline
		\end{tabular}
	}
\end{table}


\section{Analysis of Postural Instability in Parkinson's Disease Patients}
\label{sec:application}

Parkinson's disease (PD) is a neurodegenerative condition characterized by impaired motor control and various movement-related symptoms. One prominent symptom of PD is postural instability, which affects both static activities like standing upright and dynamic movements such as walking, turning, and rising from a seated position \citep{kimetal2013}. This symptom significantly affects the quality of life for individuals with PD as it increases the risk of falling, potentially leading to injuries and reduced mobility \citep{grimbergenetal2013}. Our goal is to investigate the underlying physiological mechanisms of postural regulation in PD patients and to examine the associations between postural instability and a range of clinical and behavioral outcomes. By uncovering these associations, we aim to identify contributing factors to postural instability and develop targeted interventions to enhance balance and mitigate fall risks among PD patients.

Our analysis focuses on 30-second center of pressure (COP) trajectories sampled at 25 Hz during standing with eyes closed and feet together for 29 PD patients with an average age of 68 years old. A low-pass fourth-order Butterworth with a 10 Hz cutoff frequency was applied to each COP trajectory \citep{hernandez2015correlation}. We focus on the analysis of sway length or velocity series derived from COP in AP and ML directions. In particular, let $Z_{jt}^{AP}$ and $Z_{jt}^{ML}$ denotes the COP in AP and ML directions, respectively. The sway length or velocity series is calculated as $Z_{jt} = \sqrt{(Z_{jt}^{AP}-Z_{jt-1}^{AP})^2 + (Z_{jt}^{ML}-Z_{jt-1}^{ML})^2}$ \citep{quijoux2021review}. Alongside COP trajectories, seven outcomes are also collected for each patients. These outcomes are (1) Age in years; (2) Tinetti Falls Efficacy Scale (FES), a clinically validated summary measure of fear of falling with a larger value indicating increased fear of falling \citep{tinettiFES}; (3) Activities-Specific Balance Confidence Scale (ABC), a summary measure of level of confidence in doing the activity without losing your balance with a smaller value represnting less confidence \citep{raad2013brief}; (4) TUG as the time (in second) to complete the Timed Up and Go Test \citep{nocera2013using}; (5) 4SST as the time (in second) to complete the Four Square Step Test \citep{duncan2013four}; (6) PWS as the preferred walking speed for 10 meters walk time; (7) FWS as the self-selected fast walking speed for 10 meters walk time. The mean, standard deviation, and correlation matrix of these outcomes are provided in the Supplementary Materials, revealing significant interrelationships among them. Therefore, we use the envelope-based two-stage estimator (cepENV) since it is more robust in the presence of correlation \citep{cook2013} when the sample size is not large. To aid the interpretation, we standardized the covariates. 
\begin{figure}[http]
	\centering
	\includegraphics[width=\textwidth]{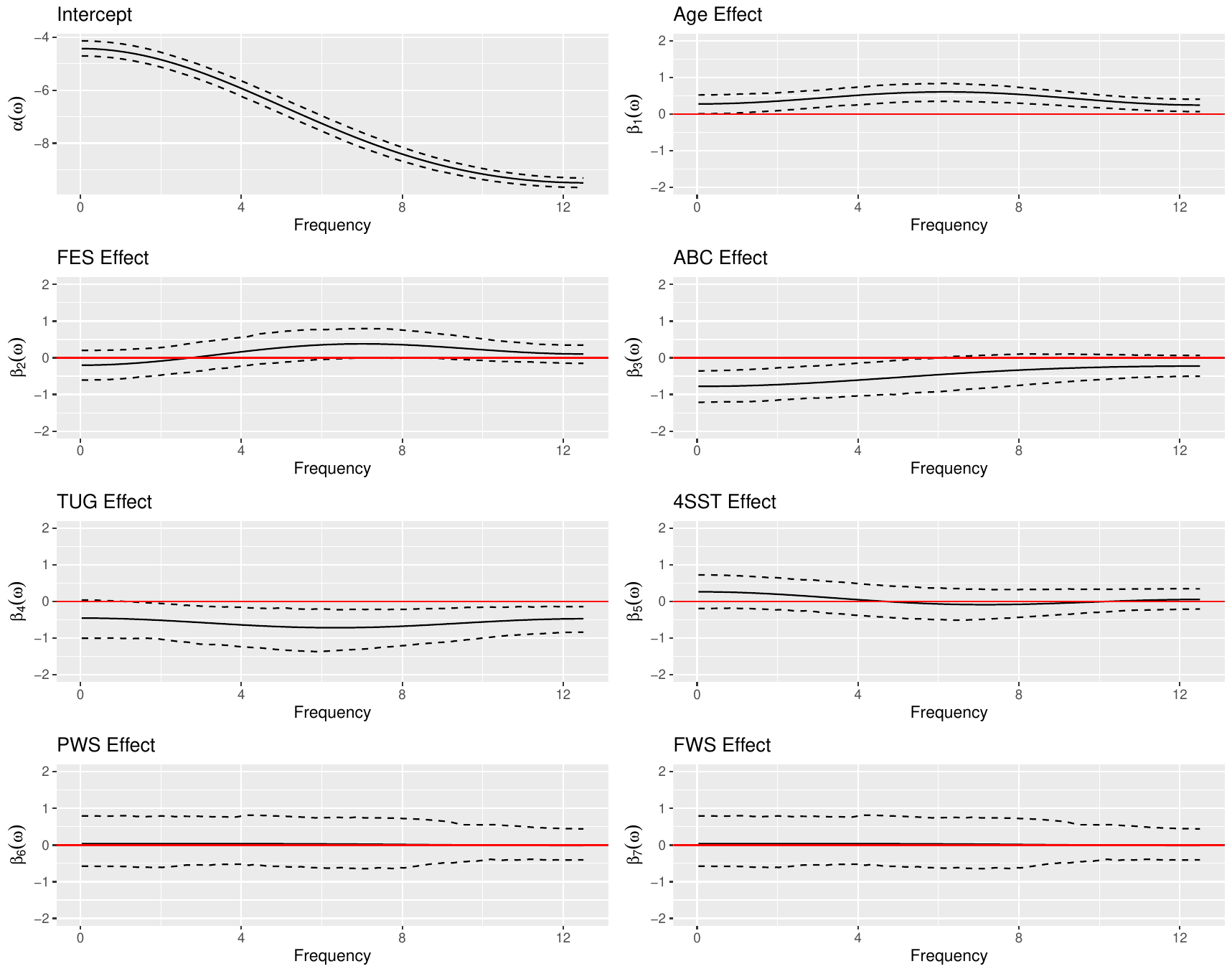}
	\caption{Estimated effect functions and the corresponding corresponding 95\% pointwise bootstrap confidence intervals from 500 bootstrap samples for the analysis of postural instability in Parkinson's disease patients. Covariates include age, Tinetti Falls Efficacy Scale (FES), Activities-Specific Balance Confidence Scale (ABC), the time (in second) to complete the Timed Up and Go Test (TUG), the time (in second) to complete the Four Square Step Test (4SST), the preferred walking speed (PWS), the self-selected walking speed (FWS).}
	\label{effect}
\end{figure}

The estimated effect functions are displayed in Figure \ref{effect}, along with the pointwise 95\% bootstrap confidence intervals based on $B = 500$ random bootstrap samples. From these plots, we can draw five important conclusions. First, the functional effects of age are significantly greater than zero across almost all frequencies, indicating that the variability of COP trajectories increases with age. The effects are strongest in the 4-8 Hz range, which aligns with previous research findings that power in the 1-10 Hz frequency band is particularly affected by age in older adults \citep{fujimoto2015effect}. Second, the FES effect function on the log-spectral is negative at low frequencies (0-2 Hz) and is significantly positive at high frequencies ($>$4 Hz). This suggests that as FES increases, the ratio of power from high frequencies to low frequencies also increases. This finding is consistent with the clinical observation that PD patients often experience stiffness and tremors in leg and trunk muscles \citep{marsden1985defects,burleighetal1995}, leading to rapid, jerky movements and higher frequency variability in COP trajectories \citep{Rocchietal2002}. Third, the functional effects of ABC are negative across almost all frequencies, indicating that the power over all frequencies decreases as ABC scores increase. This suggests that patients with higher ABC are more confident and tend to have more stable neuromuscular control.  Fourth, TUG and 4SSST, although both designed to measure postural control ability, have different effects on the log-spectra. This finding suggests that these tests may capture distinct aspects of postural control and should be further investigated in future studies to determine which test is more accurate in quantifying postural instability. Lastly, the covariates PWS and FWS are not significantly different from zero across all frequencies, indicating that self-chosen walking speeds are not good indicators of postural instability as patients tend to adapt their walking speed based on their individual conditions and comfort.

\section{Discussion}

This article introduces a flexible and computationally efficient method for analyzing the association between power spectra of replicated time series and covariates. The proposed method outperforms existing approaches in terms of estimation accuracy and computational time. However, there are some limitations and potential extensions of the proposed method that should be mentioned. First, in many studies, there may exist additional spectral variability in the multiple time series due to clustering effects \cite{krafty2011}. An important direction for future research is to extend the proposed model to incorporate possible random effects. Second, the paper explores three estimation methods, but the choice of estimator can be tailored to the specific application.   For example, if the coefficient matrix $\mathbf{B}$ is sparse with either relatively large $K$ or high-dimensional covariates, existing sparse estimation procedures for the multivariate linear model, such as those in \cite{lutz2006boosting, rothman2010sparse, wang2015joint}, could be adopted. If the power spectra are believed to be localized in a few frequencies and the covariates have sparse effects on the power spectra, sparse reduced-rank regression with various penalties could be adopted \citep{chen2012sparse,tuft2023spectra}. Moreover, for scenarios where power at specific frequencies is invariant to predictor changes and has zero regression coefficients, the so-called sparse envelope model would be appropriate \citep{su2016sparse}. If specific predictors are of particular interest, the partial envelope estimator proposed by \cite{su2011} could be a good choice. Third, the proposed method considers multiple stationary univariate time series. However, many time series, such as EEG and fMRI, are multivariate or nonstationary time series \citep{li2019, li2021h, li2024anopow}. It is of great interest to extend the proposed model to handle replicated nonstationary or multivariate time series. Finally, the proposed model utilizes the cepstral coefficients of second-order log-spectra, which is subject to some limitations. For example, oscillatory information beyond the second moment such as time-irreversibility and heavy-tail dependence cannot be accommodated. Future research could be directed to develop a cepstral representation of the copula spectral density, which inherits the robustness properties of quantile regressions and does not rely on moment assumptions \citep{kley2016, li2023robust}.

\section*{Acknowledgment}
The authors thank the referees, associate editor, and editor for providing insightful comments that greatly improved the article. The authors thank Dr. Clinton Wutzke at Gonzaga University for providing the dataset of the study on postural control in people with Parkinson's disease.

\section*{Supplementary Materials}
Supplementary materials are available online, including a pdf file that includes additional simulation results, additional results of the real data analysis, and proofs. R code for implementing the proposed method is also provided.

%
%
%

\bibliographystyle{asa}
\bibliography{mybib.bib}
\end{document}